\documentclass[conference]{IEEEtran}
\usepackage{amsmath,amssymb,amsfonts}
\usepackage{graphicx}
\usepackage{booktabs}
\usepackage{cite}
\usepackage{url}
\graphicspath{{figures/}}
\usepackage[hidelinks]{hyperref}
\usepackage{orcidlink}

\usepackage{flushend}

\newcommand{\E}{\mathbb{E}}
\newcommand{\Var}{\mathrm{Var}}

\newcommand{\rbar}{\bar{\rho}}

\begin{document}
\addtolength{\textheight}{2\baselineskip}
\bstctlcite{IEEEexample:BSTcontrol}

\title{Real-Time Edge-based Detection of Correlated AI Data-Center Load Episodes}

\author{%
\IEEEauthorblockN{%
Chandan~Chaudhary\orcidlink{0009-0002-2389-9568},
Abanish~Tiwari\orcidlink{0009-0003-0609-8571},~\emph{Student Member, IEEE},
Yansong~Pei\orcidlink{0000-0002-4647-7491},~\emph{Member, IEEE},\\
Mohammed~Ben-Idris\orcidlink{0000-0002-8731-8913},~\emph{Senior Member, IEEE},
and Joydeep~Mitra\orcidlink{0000-0001-9287-0983},~\emph{Fellow, IEEE}%
}
\IEEEauthorblockA{%
Electrical and Computer Engineering, Michigan State University, East Lansing, MI 48824, USA\\
E-mails: chaud152@msu.edu;\; tiwariab@msu.edu;\; peiyanso@msu.edu;\; benidris@msu.edu;\; mitraj@msu.edu%
}
\vspace{-3em}
}

\maketitle

\begin{abstract}
Artificial-intelligence data centers running bulk-synchronous training can impose sub-second power swings. When several facilities synchronize their training cycles, these load variations become spatially correlated and amplify the aggregate disturbance on the grid. A grid operator without access to data-center telemetry must infer this correlation from electrical measurements alone. However, the required observation time and the feasibility of detection on substation-deployable hardware remain uncharacterized. This paper develops a correlation-based detection method to classify the multi-facility operating regime from cross-facility power measurements. Analytical derivations and experimental validation show that the resulting detection confidence increases with the observation-window length at a rate governed by the load correlation time. The method is demonstrated in a real-time hardware-in-the-loop testbed, where load setpoints generated from a validated semi-Markov data-center load model are applied to an electromagnetic-transient grid simulation on a Real-Time Digital Simulator. A compact classifier built on pairwise power correlations runs on an edge device in this loop and determines whether the data-center load variations are independent or spatially correlated. The cross-facility correlation separates the independent and correlated cases across independent realizations. The held-out detection accuracy improves with the observation window, consistent with the predicted relation. A raw-waveform network fails to generalize, supporting pairwise correlation as the discriminative signal. The detector executes in real time on commodity edge hardware. A closed-loop demonstration against the running simulator tracks a regime change within one observation window.
\end{abstract}


\section{Introduction}

Hyperscale artificial-intelligence (AI) data centers train large models with bulk-synchronous parallel (BSP) protocols that alternate compute and communication phases. This alternation drives facility power between near-peak and near-idle on sub-second to second timescales~\cite{choukse2025power}. These swings fall in the natural power-system oscillation band~\cite{ko2026widearea}, and a 14.7~Hz transmission oscillation has already been traced to a single data center~\cite{Mishra2025}. United States data-center demand is on course to double within a few years~\cite{shehabi2024lbnl}, with facilities clustering into multi-gigawatt hubs. Their correlated power excursions amplify the aggregate disturbance beyond what the independent-load assumption anticipates~\cite{chaudhary2026spatial}. Reliability authorities now flag correlated loads as a front-line risk~\cite{nerc-large2025,nerc2024ltra}. The 2019 Eastern Interconnection event showed that a single source can perturb an entire interconnection~\cite{nerc2019oscillation}.

Fleet correlation therefore determines the disturbance severity. The operator, however, cannot observe the regime directly. Mitigation is predominantly source-side, through firmware power floors, workload reshaping, and grid-interactive control~\cite{choukse2025power,wang2025providing,zhou2026gridintelligent}, all of which require data-center telemetry that the operator does not hold. From the grid, the operator measures only electrical quantities and must infer the regime from them. The inference target is the joint spatial operating state across several facilities, a structure that classical single-meter monitoring cannot recover. This inference task is analogous to non-intrusive load monitoring~\cite{hart1992nilm}, which likewise disaggregates individual states from a single aggregate measurement.

The grid-side impact of AI workloads has been characterized through measured profiles, forced-oscillation models, and power-quality analyses that each represent a facility as a single source~\cite{ko2026widearea,aghadinuno2026investigation,chen2023datacenter,maheshwari2026powerquality,ginzburgganz2026technical}. These studies characterize the grid disturbances that individual AI data-center loads impose but do not assess their cross-facility detectability. Separately, spatial load correlation has been defined analytically, with its consequences traced for voltage, frequency, oscillatory stability, resource adequacy, and pre-dispatch resonance safety~\cite{chaudhary2026spatial,chaudhary2025loadmodel,chaudhary2026adequacy,chaudhary2026predispatch,chaudhary2026modal}. These studies characterize and predict spatial load correlation, but each analysis remains offline or confined to the planning stage. This paper extends that foundation to real-time detection on deployable hardware and derives the observation time the decision requires.

Detection is posed as a hypothesis test on the cross-facility power correlation, and the achievable detection confidence is shown to grow with observation time at a rate set by the natural time scale of the load fluctuations. The contributions are as follows.
\begin{itemize}
\item A detection-theoretic formulation that identifies the cross-facility correlation as the sufficient statistic for the operating regime and derives an observation-cost relation for the achievable detection performance as a function of window length.
\item A real-time three-device hardware-in-the-loop testbed in which a validated semi-Markov data-center load model~\cite{chaudhary2026hierarchical} drives an electromagnetic-transient grid simulation and the resulting grid measurements stream continuously to an edge device.
\item A compact correlation-feature detector deployed on an NVIDIA Jetson Orin Nano whose held-out accuracy follows the predicted observation-cost relation, generalizes across unseen realizations, and in closed-loop operation against the running simulator tracks a regime change within one observation window.
\end{itemize}

The remainder of this paper is organized as follows. Section~\ref{sec:model} develops the load model and the detection-theoretic formulation. Section~\ref{sec:method} describes the detector, the evaluation protocol, and the edge deployment. Section~\ref{sec:results} presents the test system and the results. Section~\ref{sec:conclusion} concludes the paper and provides direction for future research.

\section{System Model and Detection Formulation}
\label{sec:model}

This section develops the load model and the detection-theoretic formulation that establish the cross-facility power correlation as the sufficient statistic for the operating regime and bound the observation time needed to detect it reliably.

\subsection{AI Data-Center Fleet}
Consider $N$ AI data-center facilities on a transmission network. Each facility is converter-interfaced and runs bulk-synchronous training. Each facility's power follows the hierarchical semi-Markov data-center (HSM-DC) load model~\cite{chaudhary2026hierarchical}, which is fitted to measured generative-AI workload profiles~\cite{vercellino2026measurement}. The model does not depend on the host grid, so the analysis applies to any network hosting such facilities.

\subsection{Load Model}
Within one facility, power moves between a few operating phases as the training job cycles through computation, communication, and checkpointing. A semi-Markov chain over the state set $\mathcal{S}=\{\mathrm{PS,LC,BA,CI,AR}\}$ models these transitions, where the states correspond to the Peak-Surge, Light-Compute, Base, Checkpoint-Idle, and AllReduce-Dip phases of a training step. The chain has an embedded transition matrix $\mathbf{P}=[p_{ss'}]$, and each visit to state $s$ lasts a random time $\tau_s$ with mean $\mu_s$. The long-run fraction of time spent in each state is
\begin{equation}
\pi_s = \frac{\pi^{\mathrm{e}}_s\,\mu_s}{\sum_{s'}\pi^{\mathrm{e}}_{s'}\mu_{s'}},
\label{eq:stationary}
\end{equation}
where $\pi^{\mathrm{e}}$ is the stationary distribution of the embedded chain. States with longer average durations therefore occupy a larger share of the total run time. The power of facility $i$ then follows
\begin{equation}
P_i(t) = n_i\,g\!\left(s_i(t)\right) + \eta_i(t) + h_i(t),
\label{eq:facpower}
\end{equation}
where $g(s)$ is the per-node power in state $s$, $n_i$ is the active-node count, $\eta_i$ is a fast Ornstein--Uhlenbeck fluctuation that captures variability within a phase, and $h_i$ is a slow cooling term. Let $\Delta P_i(t)=P_i(t)-\E[P_i]$ be the zero-mean power deviation and $\sigma_i$ its standard deviation. Because this deviation is driven by state changes, it stays correlated over a time $\tau_c$ set by the mean phase duration, $\tau_c\!\sim\!\E_s[\mu_s]$. This load correlation time controls how quickly the regime can be detected.

\subsection{Load Correlation and the Aggregate Swing}
The spatial correlation model for AI data-center fleets is developed in~\cite{chaudhary2026spatial,chaudhary2026adequacy}. Facilities are coupled through a synchronization probability $p$: at each phase change of a reference facility, a follower copies the reference state with probability $p$ and acts independently otherwise. The pairwise correlation $\rho_{ij}$ between two facility powers rises with $p$, from $0$ under independence to a positive value under synchronization. The grid observes the aggregate $P_\Sigma=\sum_{i=1}^{N}\Delta P_i$, whose variance is
\begin{equation}
\Var(P_\Sigma) = N\sigma^2\bigl[1+(N-1)\rbar\bigr],
\label{eq:varamp}
\end{equation}
where $\sigma^2$ is the common per-facility variance and $\rbar$ is the mean pairwise correlation \cite{chaudhary2026adequacy, chaudhary2026spatial}. The bracketed factor equals $1$ when the facilities are independent and $N$ when they are fully correlated, so higher correlation directly amplifies the aggregate disturbance. This amplification factor is the quantity the operator must detect.

\subsection{Detection as a Hypothesis Test}
The operator watches $\{\Delta P_i(t)\}$ for a window of length $T$ and decides between two hypotheses,
\begin{equation}
\mathcal{H}_0\!:\rbar=0~\text{(independent)},\quad
\mathcal{H}_1\!:\rbar=\rho_s>0~\text{(correlated)}.
\label{eq:hyp}
\end{equation}
By \eqref{eq:varamp} the mean correlation $\rbar$ carries all the regime information, so the windowed estimate $\hat{\rbar}$ is the natural test statistic. The Fisher transform $z=\mathrm{atanh}(\hat{\rbar})$ yields a near-Gaussian distribution with variance $1/(N_{\mathrm{eff}}-3)$, where $N_{\mathrm{eff}}$ is the number of effectively independent samples in the window. Because the load signal stays correlated for a time $\tau_c$, this effective sample count is far smaller than the raw sample count,
\begin{equation}
N_{\mathrm{eff}}\approx \frac{T}{\tau_c}\ \ll\ T f_s .
\label{eq:neff}
\end{equation}
The two hypotheses are separated by $d=\mathrm{atanh}(\rho_s)\sqrt{N_{\mathrm{eff}}}$ standard deviations, and the test reaches $\mathrm{AUC}=\Phi(d/\sqrt{2})$, where $\Phi$ is the standard normal cumulative distribution. Substituting \eqref{eq:neff} gives the observation-cost relation,
\begin{equation}
\mathrm{AUC}(T)=\Phi\!\left(c\,\sqrt{T}\right),\qquad
c=\frac{\mathrm{atanh}(\rho_s)}{\sqrt{2\,\tau_c}} .
\label{eq:auclaw}
\end{equation}
This relation carries a practical message: detection improves with a longer observation window at a rate set by the load correlation time $\tau_c$, independently of the algorithm or sensor quality. The area under the curve tends to $0.5$ (chance level) as $T\!\to\!0$ and to $1$ as $T\!\to\!\infty$. The slower the load phases, the longer the operator must watch. 
\section{Correlation Detector and Evaluation}
\label{sec:method}

This section describes the detector, explains why input features determine whether the detector generalizes, and presents the evaluation and edge-deployment setup.

\subsection{Correlation-Feature Detector}
The system model shows that the mean cross-facility correlation $\rbar$ carries all the regime information, so the detector is built on that quantity. The three facility powers are down-sampled to $f_s=20$~Hz. Over a sliding window of $T$ seconds, the three pairwise Pearson correlations $\mathbf{r}=[\hat{\rho}_{12},\hat{\rho}_{13},\hat{\rho}_{23}]$ are formed from the power deviations. A small neural network with one hidden layer maps $\mathbf{r}$ to a binary label, either independent or correlated. The method uses a learned map instead of a fixed threshold for two reasons. Per \eqref{eq:neff}, the per-window correlation estimate is biased and its noise variance depends on its magnitude, so no single threshold is optimal across all window lengths. The same network also extends naturally to the finer classification of independent, partial, and correlated without redesign. The network is small enough to compile to an edge engine without approximation.

The input features determine whether the detector generalizes to unseen scenarios. A temporal convolutional network trained on the raw windowed measurements learns the classification task within one realization, but only by memorizing the training waveforms, which differ across realizations. On a different realization of the same regime, with the same correlation strength but different waveform trajectories, its accuracy falls to the chance level. The pairwise power correlations, by contrast, capture the joint structure across facilities~\cite{chaudhary2026spatial,chaudhary2026modal}, which defines the operating regime independently of the specific waveform trajectories. The correlation-based detector consequently transfers to unseen realizations, whereas the raw-waveform network cannot.

\subsection{Evaluation Protocol and Edge Deployment}
A deployed detector encounters realizations it never trained on, so the evaluation must measure generalization directly. The method uses leave-one-seed-out cross-validation: the detector trains on two random seeds and is evaluated on the held-out third, and this split is rotated over all three seeds so every seed appears in the test set once. Accuracy and area under the curve are averaged across the three folds. Within a seed, the independent and correlated runs share the same reference workload trajectory, so a matched pair differs only in the coupling strength. This design ensures the reported accuracy reflects genuine generalization to unseen realizations.

The entire pipeline, from the down-sampled window through pairwise correlation computation, standardization, and classification, is exported as a single computation graph and compiled to a half-precision TensorRT engine on the Jetson GPU. Because correlation is invariant to a common scale factor, a single global factor is applied inside the graph to keep the arithmetic within the half-precision range. The engine reproduces the full-precision decision exactly on the held-out samples. Inference latency is measured with the GPU clocks locked, over $10^5$ timed inferences after a warm-up period, and board power is read from the on-board monitor. The reported latency is the engine inference time only, measured without the surrounding pipeline overhead.

The same engine is also deployed in a closed loop against the running simulator. The workstation drives the RTDS and streams the load setpoints while the Jetson serves as the sole GTNETx2-SKT client. On each supervisory cycle, approximately every $50$~ms, the Jetson updates its sliding window with the newest samples, runs the fused engine, and returns the regime label over UDP. Every cycle is timestamped from the arrival of the triggering packet to the emitted decision. This timestamp gives the end-to-end loop latency.

\section{Test System Implementation and Results}
\label{sec:results}

This section presents the test system, the offline evaluation, and the closed-loop demonstration.

\subsection{Test System and Experimental Setup}
\begin{figure}[!htbp]
\centering
\vspace{-1em}
\includegraphics[trim=3.0cm 6.25cm 3.5cm 5.0cm,clip,width=\columnwidth]{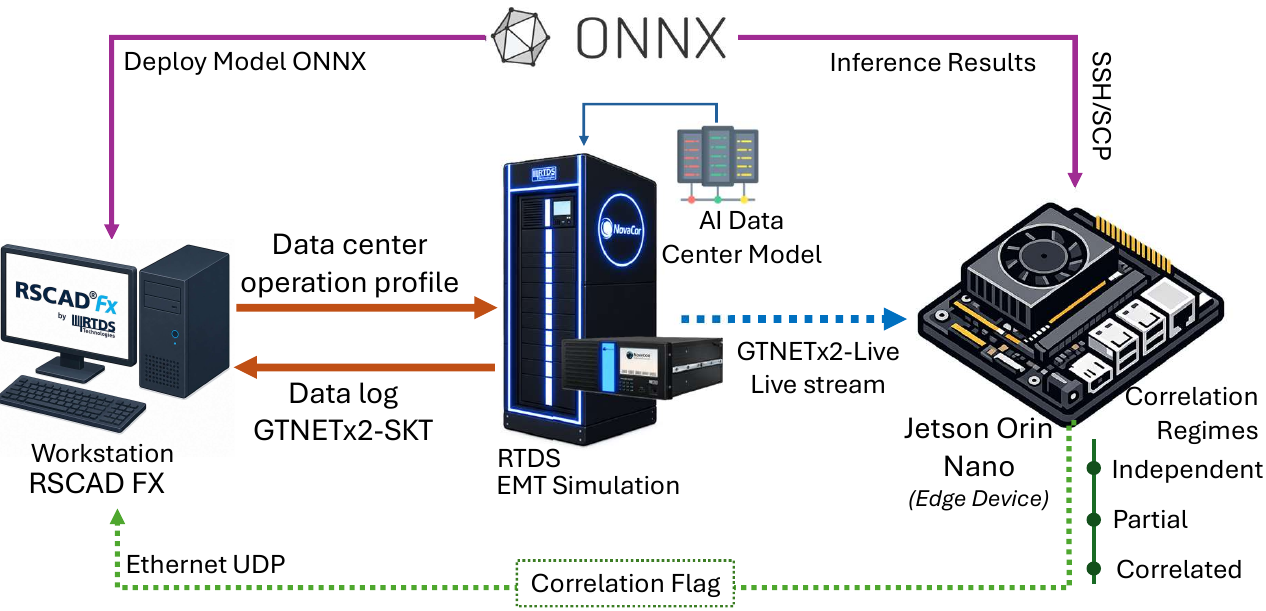}
\vspace{-1em}
\caption{Three-device real-time hardware-in-the-loop testbed. The black path is the offline data and training path; the red path is the closed-loop real-time detection path.}
\label{fig:arch}
\vspace{-1em}
\end{figure}

The three-device testbed of Fig.~\ref{fig:arch} runs the IEEE 39-bus New England network on an RTDS NovaCor~2.0 simulator at a $100~\mu$s electromagnetic-transient time step. Three AI data-center facilities connect at buses 4, 12, and 15, rated 200, 225, and 250~MW. A workstation running RSCAD~FX injects pre-generated HSM-DC load profiles as setpoints at 2~Hz. This rate resolves the second-scale BSP dynamics of \eqref{eq:facpower} but not the faster $0.2$--$3$~Hz forced-oscillation band. The RTDS streams 16 measured quantities over TCP at roughly 1~kHz: active and reactive power at each bus, system frequency, and three-phase bus currents. The NVIDIA Jetson Orin Nano~\cite{jetsonorinnano} receives this stream and hosts the correlation detector. A second path returns the edge decision over UDP. Fig.~\ref{fig:arch} distinguishes two operating modes. The offline path (solid) logs training data and deploys the compiled engine to the Jetson. The closed-loop path (dotted) connects the Jetson directly to the measurement stream and returns a regime label approximately every 50~ms. The two modes run in separate experiments because the measurement interface accepts only one TCP client at a time.

The dataset comprises nine experiments, one per regime for each of three random seeds. Each experiment logs 15~min after a 60~s warm-up. The independent regime sets the synchronization probability to zero, the partial regime uses the model-grounded coupling, and the correlated regime uses a probability of 0.45. All offline results use the leave-one-seed-out protocol. A fourth seed is reserved for the closed-loop demonstration and never appears in training or evaluation.

\subsection{The Correlation Signal}
\begin{figure}[!htbp]
\centering
\includegraphics[width=\columnwidth]{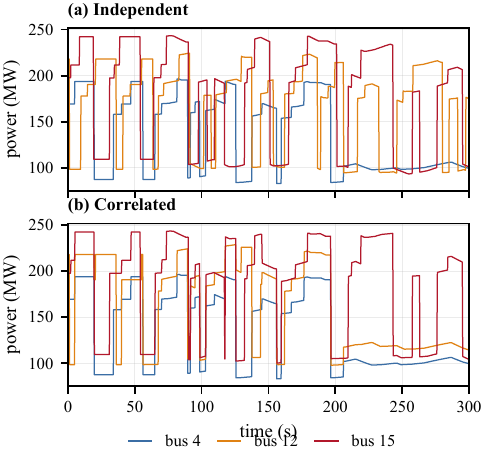}
\vspace{-2em}
\caption{Facility bus powers for one seed. (a) Independent regime. (b) Correlated regime; correlated operation aligns the three surges.}
\label{fig:ts}
\vspace{-1em}
\end{figure}

Table~\ref{tab:cmat} gives the mean pairwise correlations for each regime. The mean correlation $\hat{\rbar}$ climbs from $0.04$ under independence to $0.27$ under partial coupling and $0.43$ under full correlation. Individual pairs reach $0.54$. These values confirm the aggregate-variance amplification of \eqref{eq:varamp}~\cite{chaudhary2026spatial}. Fig.~\ref{fig:ts} shows this correlation contrast over time. Under correlation the three facility powers rise and fall together, which builds the amplified aggregate swing. Under independence they drift out of step.

\begin{table}[!htbp]
\centering
\vspace{-1.5em}
\caption{Mean Cross-Facility Power Correlation by Regime}
\label{tab:cmat}
\vspace{-1em}
\begin{tabular}{lcccc}
\toprule
Regime & $\hat{\rho}_{4,12}$ & $\hat{\rho}_{4,15}$ & $\hat{\rho}_{12,15}$ & $\hat{\rbar}$ \\
\midrule
Independent & $-0.04$ & $0.17$ & $-0.00$ & $0.04$ \\
Partial     & $0.42$  & $0.29$ & $0.11$  & $0.27$ \\
Correlated  & $0.48$  & $0.54$ & $0.28$  & $0.43$ \\
\bottomrule
\end{tabular}
\vspace{-1.5em}
\end{table}

\subsection{Detection Performance and the Observation-Cost Relation}
Table~\ref{tab:acc} reports held-out accuracy and area under the curve as a function of observation window length. At a 60~s window, the detector reaches an accuracy of 0.77 and an area under the curve of 0.84, both above the 0.667 majority-class baseline. The standard deviation across folds stays below 0.04. Performance rises monotonically with the window, as the observation-cost relation \eqref{eq:auclaw} predicts.

\begin{table}[!htbp]
\centering
\vspace{-1em}
\caption{Held-Out Binary Detection Performance (Leave-One-Seed-Out)}
\label{tab:acc}
\vspace{-1em}
\begin{tabular}{lcccc}
\toprule
Window $T$ & 3.2~s & 15~s & 30~s & 60~s \\
\midrule
Accuracy & 0.673 & 0.691 & 0.742 & 0.768 \\
AUC      & 0.706 & 0.759 & 0.774 & 0.842 \\
\bottomrule
\end{tabular}
\vspace{-1em}
\end{table}
\enlargethispage{\baselineskip}

\begin{figure}[!htbp]
\centering
\includegraphics[width=0.92\columnwidth]{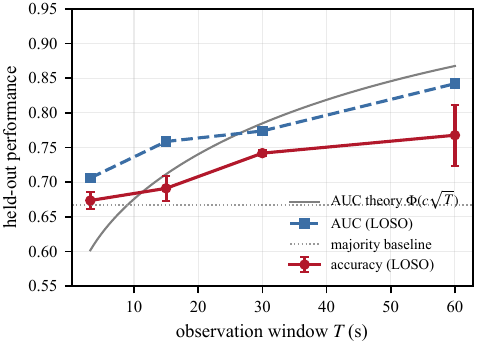}
\vspace{-1em}
\caption{Held-out detection performance against observation window. The measured area under the curve follows the predicted relation $\mathrm{AUC}$, fitted with a single parameter.}
\label{fig:acc}
\vspace{-1em}
\end{figure}

Fig.~\ref{fig:acc} overlays the measurements on the relation \eqref{eq:auclaw}. A one-parameter fit gives $c=0.14~\mathrm{s}^{-1/2}$, and the curve tracks the measured points across the full window range. With the measured correlated-regime correlation $\rho_s\approx0.43$, the fit implies a load correlation time $\tau_c\approx5$~s. This correlation time matches the mean semi-Markov sojourn of the HSM-DC model and is consistent with the 4~s intensification lead time reported in~\cite{chaudhary2026modal}. The agreement confirms that detection is limited by the observation window and the slow load dynamics. This relation also gives the operator a principled way to set the observation window for a target confidence level.

\subsection{The Cost of the Wrong Representation}
The choice of input representation determines whether detection generalizes. The raw-waveform temporal convolutional network reached $74.5\%$ accuracy when training and test windows came from one realization. Under the leave-one-seed-out protocol, accuracy fell to $0.34$, the three-class chance level, because the network memorized the training waveforms, which differ across realizations. The correlation detector keeps the performance of Table~\ref{tab:acc} under the same protocol. The contrast confirms the cross-facility correlation as the discriminative signal.

\subsection{Edge Feasibility}
Table~\ref{tab:lat} reports the inference latency of the deployed engine on the Jetson Orin Nano with the GPU clocks locked. The fused correlation-and-classifier engine reaches a median latency of $0.054$~ms at the 15~W power mode and $0.050$~ms at 25~W. The 99th-percentile latency stays at or below $0.074$~ms in both modes. Every prediction from this half-precision engine matches the full-precision model exactly. The detection cadence spans tens of seconds, so the inference time is smaller by about six orders of magnitude and the edge device is never the bottleneck.

\begin{table}[!htbp]
\centering
\vspace{-1em}
\caption{Measured Edge Inference on the Jetson Orin Nano (Clocks Locked)}
\label{tab:lat}
\vspace{-1em}
\resizebox{\columnwidth}{!}{%
\begin{tabular}{lcccc}
\toprule
Mode & Median (ms) & 99th pct.\ (ms) & Throughput (inf/s) & Power (W) \\
\midrule
15~W & 0.054 & 0.067 & 12\,659 & 6.67 \\
25~W & 0.050 & 0.074 & 12\,013 & 7.50 \\
\bottomrule
\end{tabular}%
}
\vspace{-1em}
\end{table}

\subsection{Closed-Loop Detection}
All preceding results are obtained offline, from a recorded measurement stream. This subsection closes the loop. The compiled engine runs on the Jetson while the Jetson reads the RTDS stream directly over the closed-loop path of Fig.~\ref{fig:arch}. The Jetson is the sole measurement client, and the workstation drives only the load setpoints. Every closed-loop run uses a fourth seed, a realization held out of training and evaluation.

In a fixed correlated run of five minutes, the sliding $60$~s window fills at $t=60$~s, and the edge device flips its decision to correlated at $t=60.8$~s. Across 4760 inference cycles on the roughly $50$~ms cadence, confidence keeps rising from $P(\mathrm{correlated})=0.50$ at the first flip to $0.66$ by $t=300$~s as the window accumulates more correlated signal.

\begin{figure}[!htbp]
\centering
\includegraphics[width=\columnwidth]{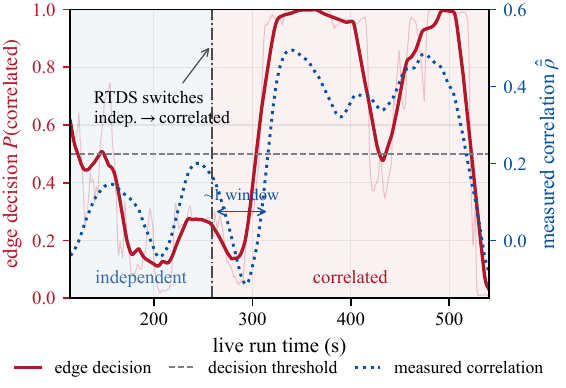}
\vspace{-1em}
\caption{Regime tracking during an independent-to-correlated transition (closed-loop run). The measured correlation $\hat{\rbar}$ (dotted) and the edge decision $P(\mathrm{correlated})$ (solid) both rise across the switch, and the decision settles within about one observation window.}
\label{fig:live}
\vspace{-1em}
\end{figure}

Fig.~\ref{fig:live} shows a second run in which the fleet transitions from independent to correlated at $t\approx259$~s. The measured correlation $\hat{\rbar}$ rises from about $0.1$ to $0.4$ across the transition. The edge decision $P(\mathrm{correlated})$ crosses the threshold and settles near $0.95$ within about $60$~s of the switch, one observation window. This settling time matches the prediction of \eqref{eq:auclaw}. The detector requires about one observation window, whether the correlation is estimated from a recording or tracked in real time as the regime shifts.

The end-to-end loop latency, measured on the fixed closed-loop run from packet arrival to the decision, is $1.16$~ms at the median and $1.50$~ms at the 99th percentile. This latency exceeds the $0.054$~ms engine-only result of Table~\ref{tab:lat} because it includes the full inference call ($0.67$~ms median, which spans host-to-device transfer, GPU execution, and device-to-host transfer) and the surrounding network and window-update overhead. Even so, the loop consumes only $1.16/50\approx2.3\%$ of the $50$~ms supervisory cadence, and the timing margin against the seconds-scale correlation time $\tau_c$ spans several orders of magnitude.

\subsection{Discussion and Limitations}
These results turn the qualitative concern over correlated AI loads into a measured trade-off between detection confidence and observation time. The cross-facility correlation is the discriminative signal, and the observation-cost relation prescribes how long the operator must observe. The closed-loop test confirms that a substation-grade edge device can resolve the regime from the measurement stream. Four limitations bound the study. First, the correlated regime is produced by a modeled synchronization probability and has not been confirmed against a measured multi-facility grid event. Second, the detector is evaluated at the facility buses, not at a remote or aggregated bus where the correlation signature is weaker. Third, the held-out estimate averages over three seeds, which limits statistical precision. The closed-loop demonstration draws on a single additional seed. Fourth, the Jetson runs a soft real-time Linux stack, so the reported latency is a statistical measure with no hard real-time guarantee. The 2~Hz load injection resolves only the seconds-scale surge studied here, and the faster $0.2$--$3$~Hz forced-oscillation band is left for future work.

\section{Conclusion}
\label{sec:conclusion}

This paper established that correlated AI data-center load episodes can be detected in real time from grid measurements alone, without access to data-center telemetry. The cross-facility power correlation is the sufficient signal that separates the operating regimes, and the observation-cost relation governs how quickly any operator can achieve a given detection confidence, regardless of the classifier used. The fundamental limit is the load correlation time, which sets the rate at which confidence grows with observation window length. A hardware-in-the-loop demonstration confirms these properties in real-time operation. A commodity edge device closes the detection loop against a running grid simulator, resolves a regime change within one observation window, and executes within a fraction of the supervisory cadence.

Future work will extend the observation to a remote or aggregated bus, increase the injection bandwidth to cover the faster forced-oscillation band, and close the control loop so the edge decision drives a corrective reserve or curtailment action. Together, these extensions would advance the method from passive detection to autonomous grid-protective control on substation hardware.

\section*{Acknowledgment}
This research was supported in part by the MSU Research Foundation and in part by the U.S. National Science Foundation under Grant No.~2408615.

\bibliographystyle{IEEEtran}
\bibliography{references}

@IEEEtranBSTCTL{IEEEexample:BSTcontrol,
  CTLdash_repeated_names = "no"
}

@article{choukse2025power,
  author  = {Choukse, Esha and Warrier, Brijesh and Heath, Scot and Belmont, Luz and Zhao, April and Khan, Hassan Ali and Harry, Brian and Kappel, Matthew and Hewett, Russell J. and Datta, Kushal and others},
  title   = {Power Stabilization for {AI} Training Datacenters},
  journal = {arXiv preprint arXiv:2508.14318},
  year    = {2025}
}

@article{ko2026widearea,
  author  = {Ko, Min-Seung and Zhu, Hao},
  title   = {Wide-Area Power System Oscillations from Large-Scale {AI} Workloads},
  journal = {{IEEE} Transactions on Power Systems},
  year    = {2026},
  pages   = {1--14},
  doi     = {10.1109/TPWRS.2026.3685506}
}

@article{Mishra2025,
  author  = {Mishra, Chetan and Vanfretti, Luigi and De La Ree, Jaime and Purcell, T. J. and Jones, Kevin D.},
  title   = {Understanding the Inception of 14.7 {Hz} Oscillations Emerging from a Data Center},
  journal = {Sustainable Energy, Grids and Networks},
  volume  = {43},
  pages   = {101735},
  year    = {2025},
  doi     = {10.1016/j.segan.2025.101735}
}

@inproceedings{aghadinuno2026investigation,
  author    = {Aghadinuno, Chinwe and Ahmed, Sara and Alamaniotis, Miltos and Wang, Bin and Gatsis, Nikolaos},
  title     = {Investigation of {AI} Data Center Load Impact on Power System Frequency Using Real-World Datasets},
  booktitle = {2026 {IEEE} Green Technologies Conference (GreenTech)},
  year      = {2026},
  pages     = {1--6},
  doi       = {10.1109/GreenTech68285.2026.11471548}
}

@article{chen2023datacenter,
  author  = {Chen, Yenan and Shi, Keyan and Chen, Min and Xu, Dehong},
  title   = {Data Center Power Supply Systems: From Grid Edge to Point-of-Load},
  journal = {{IEEE} Journal of Emerging and Selected Topics in Power Electronics},
  volume  = {11},
  number  = {3},
  pages   = {2441--2456},
  year    = {2023},
  doi     = {10.1109/JESTPE.2022.3229063}
}

@inproceedings{maheshwari2026powerquality,
  author    = {Maheshwari, Lakshman and Dai, Hang and Sarlioglu, Bulent and Chakraborty, Rahul and Xia, Yu and Harper, Mario},
  title     = {{AI} Data Centers: A Review of Power Quality Challenges and Mitigation Strategies},
  booktitle = {2026 {IEEE} Power and Energy Conference at {Illinois} (PECI)},
  year      = {2026},
  pages     = {1--6},
  doi       = {10.1109/PECI70026.2026.11516457}
}

@article{ginzburgganz2026technical,
  author  = {Ginzburg-Ganz, Elinor and Lifshits, Pavel and Machlev, Ram and Belikov, Juri and Krieger, Ziv and Levron, Yoash},
  title   = {Technical Challenges of {AI} Data Center Integration into Power Grids---{A} Survey},
  journal = {Energies},
  volume  = {19},
  number  = {1},
  pages   = {137},
  year    = {2026},
  doi     = {10.3390/en19010137}
}

@article{wang2025providing,
  author  = {Wang, Yi and Guo, Qinglai and Chen, Min},
  title   = {Providing Load Flexibility by Reshaping Power Profiles of Large Language Model Workloads},
  journal = {Advances in Applied Energy},
  volume  = {19},
  pages   = {100232},
  year    = {2025},
  doi     = {10.1016/j.adapen.2025.100232}
}

@article{zhou2026gridintelligent,
  author  = {Zhou, Yihong and Morstyn, Thomas},
  title   = {Grid-Intelligent {AI} Data Centres for Primary Response},
  journal = {{IEEE} Transactions on Industry Applications},
  year    = {2026},
  pages   = {1--14},
  doi     = {10.1109/TIA.2026.3678552},
  note    = {Early access}
}

@misc{vercellino2026measurement,
  author        = {Vercellino, Roberto and Willard, Jared and Campos, Gustavo and Pereira, Weslley da Silva and Hull, Olivia and Selensky, Matthew and Mueller, Juliane},
  title         = {Measurement of Generative {AI} Workload Power Profiles for Whole-Facility Data Center Infrastructure Planning},
  year          = {2026},
  eprint        = {2604.07345},
  archivePrefix = {arXiv},
  primaryClass  = {eess.SY},
  doi           = {10.48550/arXiv.2604.07345}
}

@techreport{nerc2024ltra,
  author      = {{North American Electric Reliability Corporation}},
  title       = {{2024 Long-Term Reliability Assessment}},
  institution = {NERC},
  address     = {Atlanta, GA},
  year        = {2024},
  month       = dec
}

@techreport{nerc-large2025,
  author      = {{North American Electric Reliability Corporation}},
  title       = {{Characteristics and Risks of Emerging Large Loads}},
  institution = {NERC},
  type        = {Large Loads Task Force White Paper},
  year        = {2025},
  month       = jul
}

@techreport{nerc2019oscillation,
  author      = {{North American Electric Reliability Corporation}},
  title       = {{January 11, 2019 Eastern Interconnection Forced Oscillation Event}},
  institution = {NERC},
  year        = {2019}
}

@techreport{shehabi2024lbnl,
  author      = {Shehabi, Arman and Smith, Sarah Josephine and Hubbard, Alex and Newkirk, Alexander and Lei, Nuoa and Siddik, Md AbuBakar and Holecek, Billie and Koomey, Jonathan G. and Masanet, Eric R. and Sartor, Dale A.},
  title       = {{United States Data Center Energy Usage Report}},
  institution = {Lawrence Berkeley National Laboratory},
  year        = {2024},
  month       = dec,
  doi         = {10.71468/P1WC7Q}
}

@article{hart1992nilm,
  author  = {Hart, George W.},
  title   = {Nonintrusive Appliance Load Monitoring},
  journal = {Proceedings of the {IEEE}},
  volume  = {80},
  number  = {12},
  pages   = {1870--1891},
  year    = {1992}
}

@inproceedings{chaudhary2026spatial,
  author    = {Chaudhary, Chandan and Abdelkader, Alaaeldein and Pei, Yansong and Benidris, Mohammed and Mitra, Joydeep},
  title     = {Spatial Load Correlation in {AI} Data-Center-Dominated Power Systems},
  booktitle = {2026 {IEEE} Power \& Energy Society General Meeting (PES GM)},
  address   = {Montr{\'e}al, QC, Canada},
  month     = jul,
  year      = {2026}
}

@inproceedings{chaudhary2026modal,
  author    = {Chaudhary, Chandan and Murillo, Michael and Benidris, Mohammed and Mitra, Joydeep and Pandit, Dilip and Bera, Atri},
  title     = {Modal Analysis of Spatial Load Correlation in {AI} Data Center-Dominated Power Systems},
  booktitle = {2026 {IEEE} Int.\ Conf.\ Smart Energy Systems and Technologies (SEST)},
  month     = sep,
  year      = {2026}
}

@inproceedings{chaudhary2026adequacy,
  author    = {Chaudhary, Chandan and Abdelkader, Alaaeldein and Benidris, Mohammed and Mitra, Joydeep},
  title     = {Resource Adequacy Risk in Correlated Large Loads},
  booktitle = {2026 {IEEE} Int.\ Conf.\ Probabilistic Methods Applied to Power Systems (PMAPS)},
  address   = {Salt Lake City, UT, USA},
  month     = sep,
  year      = {2026}
}

@inproceedings{chaudhary2026predispatch,
  author    = {Chaudhary, Chandan and Tiwari, Abanish and Pei, Yansong and Benidris, Mohammed and Mitra, Joydeep},
  title     = {A Pre-Dispatch Resonance Safety Criterion for {AI} Training Clusters},
  booktitle = {2026 North American Power Symposium (NAPS)},
  address   = {Houghton, MI, USA},
  month     = oct,
  year      = {2026}
}

@inproceedings{chaudhary2025loadmodel,
  author    = {Chaudhary, Chandan and Abdelkader, Alaaeldein and Egan, Michael and Udren, Eric and Benidris, Mohammed and Mitra, Joydeep},
  title     = {Impact of Data Center Load Modeling on Power System Stability},
  booktitle = {CIGRE US Grid of the Future Symposium},
  address   = {Denver, CO, USA},
  month     = nov,
  year      = {2025}
}

@inproceedings{chaudhary2026hierarchical,
  author    = {Chaudhary, Chandan and Bera, Atri and Pandit, Dilip and Newlun, Cody and Ben-Idris, Mohammed and Mitra, Joydeep},
  title     = {A Hierarchical Semi-{Markov} Load Model for {AI} Data Centers Coupling Job Scheduling with Bulk-Synchronous-Parallel Power Dynamics},
  booktitle = {Proceedings of the CIGRE Grid of the Future Symposium},
  year      = {2026},
  month     = oct,
  note      = {Submitted, under review}
}

@misc{jetsonorinnano,
  author = {{NVIDIA}},
  title  = {{Jetson Orin Nano Series Modules Data Sheet}},
  year   = {2025}
}

\end{document}